\documentclass{article}

\usepackage{PRIMEarxiv}

\usepackage[utf8]{inputenc} 
\usepackage[T1]{fontenc}    
\usepackage{hyperref}       
\usepackage{url}            
\usepackage{booktabs}       
\usepackage{amsfonts}       
\usepackage{nicefrac}       
\usepackage{microtype}      
\usepackage{lipsum}
\usepackage{fancyhdr}       
\usepackage{graphicx}       
\graphicspath{{media/}}     

\usepackage{tikz}
\usetikzlibrary{arrows,automata,positioning}

\usepackage{mathtools}
\usepackage{xspace}

\usepackage{hyperref}

\usepackage[utf8]{inputenc} 
\usepackage[T1]{fontenc}    
\usepackage{hyperref}       
\usepackage{url}            
\usepackage{booktabs}       
\usepackage{amsfonts}       
\usepackage{nicefrac}       
\usepackage{microtype}      
\usepackage{xcolor}         
\usepackage{answers}

\usepackage{caption}
\usepackage[ruled,vlined]{algorithm2e}

\usepackage{bbm}
\usepackage{tikz}
\usetikzlibrary{arrows,automata,positioning}

\usepackage[ruled]{algorithm2e}
\usepackage{algpseudocode}

\title{Variance Reduction of Resampling for Sequential Monte Carlo
}

\author{
  Xiongming Dai  \\
  Division of Computer Science and Engineering \\
  Louisiana State University \\
  Baton Rouge,LA70803, USA\\
  \texttt{\{xdai2\}@email} \\
   \And
  Gerald Baumgartner \\
  Division of Computer Science and Engineering \\
  Louisiana State University \\
  Baton Rouge,LA70803, USA\\
  \texttt{\{gb\}@email} \\
}

\begin{document}
\maketitle

\begin{abstract}

 A resampling scheme provides a way to switch low-weight particles for sequential Monte Carlo with higher-weight particles
  representing the objective distribution. The less the variance of the weight distribution is, the more concentrated the
  effective particles are, and the quicker and more accurate it is to approximate the hidden Markov model, especially for
  the nonlinear case. We propose a repetitive deterministic domain with median ergodicity for resampling and have achieved the lowest
  variances compared to the other resampling methods. As the size of the deterministic domain $M\ll N$ (the size of population), given a feasible size of particles, our algorithm is faster than the state of the art, which is verified by theoretical deduction and
  experiments of a hidden Markov model in both the linear and non-linear cases.

\end{abstract}


\section{Introduction}

Sequential Monte Carlo (SMC) or Particle Filter~\cite{gordon1993novel} 
is a set of Monte Carlo methods for solving
nonlinear state-space models given noisy partial observations, which are widely used in signal and image processing
\cite{sarkka2007rao}, stock analysis \cite{casarin2006business,flury2011bayesian}, or robotics \cite{fox2001kld}. 
It updates the predictions recursively by samples composed of weighted particles to infer the posterior probability
density. While the particles will be impoverished as the sample forwards recursively, it can be mitigated by resampling where the negligible weight particles will be replaced by other particles with higher weights \cite{doucet2000sequential}.

In the literature, several resampling methods and corresponding theoretical analysis
\cite{kunsch2005recursive,chopin2004central,douc2007limit,gilks2001following} can be found.
The frequently used algorithms are residual resampling~\cite{liu1998sequential}, multinomial
resampling~\cite{gordon1993novel}, stratified resampling \cite{smith2013sequential},
and systematic resampling \cite{kitagawa1996monte,arulampalam2002tutorial}.
A justified decision regarding which resampling
strategies to use might result in a reduction of the overall computation effort and high accuracy of the estimations of the
objective. However, for resampling, most of these strategies traverse repetitively from the original population, the negligible
weight particles fail to be discarded completely, although the diversity of the particle reserve, causes unnecessary
computational load and affects the accuracy of estimations of the posterior distribution.
From the perspective of complexity and
variance reduction with promising estimation, we propose a repetitive deterministic domain ergodicity strategy, where more
concentrated and effective particles are drawn to approximate the objective. Our proposal can be widely used in large-sample
approximations.

In this paper, we concentrate on the analysis of the importance sample resamplings built-in SMC for the hidden Markov model.
In Section 2, we present a brief introduction to SMC\@. Here, a brief introduction to the hidden Markov model and the
sequential importance sampling method will be given. Our method will be introduced in Section 3, where we introduce the origin of our method, and how to implement each step in detail, and then the theoretical asymptotic behavior of approximations using our method is provided. The practical experiments
will be validated by Section 4, where performance and complexity analysis are presented.
The summary of our contributions
is outlined in Section 5.

\section{Resampling in SMC for Hidden Markov Model}

Consider the state-space model, which is also known as a hidden Markov model, described by 

\begin{equation}
  X_0\sim \mu(X_0), \ \ \ \
  X_t\mid X_{t-1} \sim f(X_t \mid X_{t-1}), \ \ \ \
  Y_t\mid X_t \sim g(y_t \mid X_t).
\end{equation}

 The initial state ${X_0}$ follows probability density distribution $\mu(X_0)$, $X_t,(t=1,2,...n)$ is a latent variable to be observed, the measurements ${Y_t}$ are assumed to be conditionally independent given ${X_t}$, the most objective is to estimate ${X_t}$.
 
The recursive Bayesian estimation can be used and it is described as:\\
(a) Prediction
\begin{equation}\label{2}
 \pi(X_t\mid y_{1:t-1})=\int f(X_t\mid X_{t-1})\pi(X_{t-1}\mid y_{1:t-1})dX_{t-1}
\end{equation}
(b) Update
\begin{equation}\label{3}
 \pi(X_t\mid y_{1:t})=\frac{g(y_t\mid X_t)\pi(X_t\mid y_{1:t-1})}{\int g(y_t\mid X_t)\pi(X_t\mid y_{1:t-1})dX_t}
\end{equation}
 
 From~\eqref{2} and~\eqref{3} the integral part is unreachable, especially,
 for high-dimensional factors involved, we fail to get the close form of
 $\pi(X_t\mid y_{1:t})$~\cite{sarkka2013bayesian,doucet2009tutorial}. 
 
 Sequential Monte Carlo is a recursive algorithm where a cloud of particles is propagated to approximate the posterior distribution $\pi (X_{0:t} \mid y_{1:t})$. Here, we describe a general algorithm that generates at time $t$, $N$ particles  $\left \{{X_{0:t}^{(i)}}  \right \}_{i=1}^N$ with the corresponding empirical measure   $ \hat{\pi }(X_{0:t} \mid y_{1:t})= \sum_{i=1}^{N} w_t^i\delta _{X_{0:t}}^{(i)}(dX_{0:t})$%
 , a discrete weighted approximation of the true posterior $\pi (X_{0:t} \mid y_{1:t})$, $\delta _{X_{0:t}}^{(i)}(dX_{0:t})$ denotes the delta-Dirac mass located at $X_t$, $dX_{0:t}$ equals to $X_{0:t}-X_{0:t}^i$. The particles are drawn recursively using the observation obtained at time $t$ and the set of particles  $\left \{{X_{0:t-1}^{(i)}}  \right \}_{i=1}^N$ drawn at time ${t-1}$, accordingly, where $\hat{\pi} (X_{0:t-1} \mid y_{1:t-1})\approx {\pi} (X_{0:t-1} \mid y_{1:t-1})$. The weights are normalized using the principle of importance sampling such that $\sum_{i=1}^{N}w_t^i=1$. If the samples $X_{0:t}^i$ are drawn from an importance density $q(X_{0:t}^i\mid y_{1:t})$, we have
 
 \begin{equation}\label{equal:311}
     w_t^i\propto \frac{{\pi} (X_{0:t}^i \mid y_{1:t})}{q(X_{0:t}^i\mid y_{1:t})}
 \end{equation}
 
 Suppose at time step $t-1$, we have existed samples to approximate the current posterior distribution ${\pi }(X_{0:t-1} \mid y_{1:t-1})$, if we get a new observation $y_t$ at time $t$, a recursive approximation to ${\pi }(X_{0:t} \mid y_{1:t})$ with a new set of samples can be obtained by importance sampling, the corresponding factorization \cite {arulampalam2002tutorial} is described by
 
  \begin{equation}\label{equal:312}
     q\left ( X_{0:t} \mid y_{1:t}\right ):= q(X_t \mid X_{0:t-1},y_{1:t})q(X_{0:t-1} \mid y_{1:t-1})
 \end{equation}
 
 Then, we can get the new samples $X_{0:t}^i\sim q(X_{0:t}\mid y_{1:t})$ by propagating each of the existing samples $X_{0:t-1}^i\sim q(X_{0:t-1}\mid y_{t-1})$ with the new state $X_t^i\sim q(X_t\mid X_{0:t-1}, y_t)$. To derive the weight update equation, we follow the ergodic Markov chain properties of the model, the full posterior distribution $\pi (X_{0:t} \mid y_{1:t})$ can be written recursively in terms of $\pi (X_{0:t-1} \mid y_{1:t-1})$, $g(y_t \mid X_t)$ and $f(X_t \mid X_{t-1})$  \cite{arulampalam2002tutorial}:

 \begin{equation}\label{equal:314}
    \pi (X_{0:t} \mid y_{1:t}) = \frac{p(y_t \mid X_{0:t},Y_{1:t-1})p(X_{0:t} \mid y_{1:t-1})}{p(y_t \mid y_{1:t-1})}\Rightarrow 
 {\pi }_{1:t}\propto g(y_t \mid x_t)f(x_t \mid x_{t-1})\pi_{1:t-1}
 \end{equation}
 
Where ${\pi }_{1:t}$ is short for $\pi (X_{0:t} \mid y_{1:t})$.
By substituting ~\eqref{equal:312} and ~\eqref{equal:314} into ~\eqref{equal:311}, we have

\begin{equation} \label{eq1}
w_t^i \propto \frac{g(y_t \mid X_t^i)f(X_t^i \mid X_{t-1}^i)p(X_{0:t-1}^i \mid y_{1:t-1})}{q(X_t^i\mid X_{0:t-1}^i,Y_{1:t})q(X_{0:t-1}^i\mid Y_{1:t-1})}  = w_{t-1}^i\frac{g(y_t \mid X_t^i)f(X_t^i \mid X_{t-1}^i)}{q(X_t^i\mid X_{0:t-1}^i,y_{1:t})}\\
\end{equation}

We assume the state $X_t$ is ergodic Markovian, thus, $q(X_t^i\mid X_{0:t-1}^i,y_{1:t})=q(X_t^i\mid X_{t-1}^i,y_t)$, from this point, we only need to store the $X_{t}^i$, and obtain the thinning recursively update weight formula \cite{gordon2004beyond}: 

 \begin{equation}
 \label{increase}
     w_t^i \propto w_{t-1}^i\frac{g(y_t \mid x_t^i)f(x_t^i \mid x_{t-1}^i)}{q(x_t^i\mid x_{t-1}^i,y_t)}
 \end{equation}
The corresponding empirical posterior filtered density $\pi (X_{t} \mid y_{1:t})$ can be approximated as

 \begin{equation}
\hat{\pi} (X_{t} \mid y_{1:t})= \sum_{i=1}^{N} w_t^i\delta _{X_{t}}^{(i)}(dX_t)
 \end{equation}

It can be shown that as $N\rightarrow \infty $, $\hat{\pi} (X_{t} \mid y_{1:t})$ converges to ${\pi}_t={\pi} (X_{t} \mid y_{1:t})$.

Ideally, the importance density function should be the posterior distribution itself, ${\pi} (X_{0:t} \mid y_{1:t})$. While the variance of importance weights increases over time, which will decrease the accuracy and lead to the degeneracy that some particles make up negligible normalized weights. 
The brute force approach to reducing the effect of degeneracy is to increase $N$ as large as possible. However, as the size of the sample increases, the computation of the recursive step will also be exponentially costly. Generally, we can try two ways to improve: (I) suitable importance density sampling; (2) resampling the weights. Here we focus on the latter.  A suitable measure of the degeneracy of an algorithm is the effective sample size $N_{eff}$ introduced in \cite {arulampalam2002tutorial}:$ N_{eff}=\frac{N}{1+Var(w_t^{*i})}$,
$w_t^{*i}=\frac{\pi(X_t^i\mid y_{1:t})}{q(X_t^i\mid X_{t-1}^i,y_t)}$, while the close solution is unreachable, it could be approximated \cite{liu2008monte} by $\hat{N}_{eff}=\frac{1}{\sum_{i=1}^{N}(w_t^i)^2}$
. If the weights are uniform, $w_t^i=\frac{1}{N}$ for each particle $i=1,2,...,N$, $N_{eff}=N$; If there exists the unique particle, whose weight is $1$, the remaining are zero, $N_{eff}=1$. Hence, small $N_{eff}$ easily lead to a severe degeneracy \cite{gordon2004beyond}. 
We use $\hat{N}_{eff}$ as an indicator to measure the condition of resampling for our experiments in section 4.

We will introduce our proposal based on the repetitive deterministic domain traverse in the next section.

\section{Repetitive Deterministic Domain with Median Ergodicity Resampling}
 
 \subsection{Multinomial Sampling}

A Multinomial distribution provides a flexible framework with parameters $p_i,i=1,...,k$ and $N$, to measure the probability that each class $i \in {1,...,k}$ has been sampled $N_i$ times over $N$ categorical independent tests. It can be used to resample the location in our proposal in two steps. Firstly, we obtain the samples from a uniform generator $u^i\sim U(0,1],i=1,...,N$; secondly, we evaluate the index $j$ of samples with the generalized inverse rule, if the cumulative sum of samples  $\sum_{i=1}^{j}w_i$ larger or equal to $u^i$, this index $j$ will be labeled, then the corresponding sample $w_i$ will be resampled, this event can be mathematically termed as  $g(w_i)=\mathbb{I}_{w_i=w_j}$.

\subsection{Deterministic Domain Construction}

The population of weights is divided into two parts. The first part is the weights, larger than the average $\frac{1}{N}$, they are considered as the candidate firstly to be sampled, we keep $r_i=\left \lfloor N \hat{w}_t^i \right \rfloor$ replicates of $\hat{w}_t^i$ for each $i$, where $\hat{w}_t^i$ is the renormalized unit. 
$r_i$ will be filtered one by one from the population, and the corresponding tag $j$ will be saved into an array. We find, this part also follows the multinomial distribution ${W}^i \sim \hbox{\it Multinomial}(M;{\hat{w}^1},...{\hat{w}^M})$, We extract the samples from the population with the rule of multinomial sampling shown in section 3.1. This step is the first layer of the traverse from the population, we achieve the first subset, then, we renormalized the weights in the subset, and traverse again to differentiate the larger weights and other units, until we get the feasible size of the set to be considered as the potential deterministic domain.



\begin{tikzpicture}[shorten >=1pt,node distance=2cm,auto,roundnode/.style={circle, draw=green!60, fill=green!5, very thick, minimum size=7mm},
squarednode/.style={rectangle, draw=red!60, fill=red!5, very thick, minimum size=5mm}]
  \tikzstyle{every state}=[fill={rgb:black,1;white,10}]

  \node[state]   (s_{11})                      {$s_{11}$};
  \node[state] (s_{12}) [below of=s_{11}]  {$s_{12}$};
  \node[state]           (s_{1N}) [below of=s_{12}]     {$s_{1N}$};
  

\node[draw] at (0,1) {$t=1$};
\node[state] at (2.0,-2) (s_{22}){$s_{22}$};
  \node[state]           (s_{21}) [above of=s_{22}]     {$s_{21}$};
  \node[state]           (s_{2N}) [below of=s_{22}]     {$s_{2N}$};
\node[draw] at (2.0,1) {$t=2$};
\node[draw] at (4.5,1) {$t=T$};

\node[state] at (4.5,-2) (s_{T2}){$s_{T2}$};
  \node[state]           (s_{T1}) [above of=s_{T2}]     {$s_{T1}$};
  \node[state]           (s_{TN}) [below of=s_{T2}]     {$s_{TN}$};
    {...};
\node[text width=3cm] at (4.40,-2) 
    {$\cdots$};

\node[text width=3cm] at (0.0,-2.7) {$\vdots$}; 
\node[text width=3cm] at (2.0,-2.7) {$\vdots$};
\node[text width=3cm] at (4.50,-2.7) {$\vdots$};

  \path[->]
  (s_{11})   edge[bend left]              node {} (s_{12})
        edge[bend left]              node {} (s_{1N})
        edge [loop left]  node {} (   )
  (s_{12}) edge [loop left]  node {} (   )
        edge [bend left]  node {} (s_{1N})
        edge [bend left]  node {} (s_{11})
  (s_{1N}) edge [loop left]  node {} (   )
        edge [bend left]  node {} (s_{12})
         edge [bend left]  node {} (s_{11})
  (s_{21})   edge[bend left]              node {} (s_{22})
        edge[bend left]              node {} (s_{2N})
        edge [loop left]  node {} (   )
  (s_{22}) edge [loop left]  node {} (   )
        edge [bend left]  node {} (s_{2N})
        edge [bend left]  node {} (s_{21})
  (s_{2N}) edge [loop left]  node {} (   )
        edge [bend left]  node {} (s_{22})
         edge [bend left]  node {} (s_{21})
         
  (s_{T1})   edge[bend left]              node {} (s_{T2})
        edge[bend left]              node {} (s_{TN})
        edge [loop left]  node {} (   )
  (s_{T2}) edge [loop left]  node {} (   )
        edge [bend left]  node {} (s_{TN})
        edge [bend left]  node {} (s_{T1})
  (s_{TN}) edge [loop left]  node {} (   )
        edge [bend left]  node {} (s_{T2})
         edge [bend left]  node {} (s_{T1});
\end{tikzpicture}

We define the integer part event, $g(\hat{w}_i)=\mathbb{I}_{\hat{w}_i=\hat{w}_j}$, similarly for the following repetitive part, $\bar{g}(\bar{w}_i)=\mathbb{I}_{\bar{w}_i=\bar{w}_j}$. We count the units involved in the occurrence of the event $g(\hat{w}_i)$ and $\bar{g}(\bar{w}_i)$, then extract these units based on the tags $j$, which forms the final deterministic domain.

\subsection{Repetitive Ergodicity in Deterministic Domain with Median Schema}

Our goal is to retract and retain units with large weights, while the remaining ones with low weights can be effectively replaced in the populations. We set the desired number of resampled units as the size of populations under the premise of ensuring unit diversity as much as possible.

We normalized all the units to keep the same scaled level for comparison, after that, the units with higher weights above the average level will appear as real integers (larger than zero) by $\hbox{\it Ns} = \hbox{\it floor}(N.* w)$, the remaining will be filtered to zero. This is the prerequisite for the deterministic domain construction. In $\hbox{\it Ns}$ subset, there exist multiple categorical units, that follow the multinomial distribution. We sample these termed large units with two loops, the outer loop is to bypass the index of the unit zero, and the inner loop is to traverse and sample the subset where different large units distribute, there more large weights will be sampled multiple times.

The last procedure is to repetitively traverse in the deterministic domain, where each unit will be renormalized and the corresponding cumulative summation is used to find the index of the unit with the rule of the inverse cumulative distribution function. Each desired unit will be drawn by the multinomial sampler to rejuvenate the population recursively. The complexity
of our method is $\mathcal{O}(M)$. As the size of the deterministic domain $M\ll N$ (the size of population), given a feasible size of particles,
our algorithm is faster than the state of the art. The total implement schema is shown in Algorithm~\ref{algo:event}.

\begin{algorithm}[t]
	\caption{Repetitive Deterministic Domain Traversal Resampling}
	\label{algo:event}
	\KwIn{The input weight sequence:$w$; the desired number of resampled particles: $N$}
	\KwOut{The resampled tag of the weight sequence $tag$}
	\If{nargin==1}{
	
				 $N \leftarrow  length(w)$\tcp*{Desired  size} 
			}{}
	$M \leftarrow length(w)$;\\
    $w \leftarrow w / sum(w)$\tcp*{Normalization}
    $tag \leftarrow zeros(1, N)$;\\
    $Ns \leftarrow floor(N.* w)$\tcp*{Integer parts}
    $R \leftarrow sum(Ns)$;\\
    $i \leftarrow 1$\tcp*{Extract deterministic part}
    $j \leftarrow 0$;\\
	\While{j $<$ M}{	
	$j \leftarrow j+1$\\
	$count \leftarrow 1$\\
	    \While{count $<=$ Ns(j)}{
	    $tag(i) \leftarrow j$;\\  
    $i \leftarrow i + 1$;\\ 
    $count \leftarrow count + 1$;\\
	    }
	    }
    $[Wsorted,I] = sort(w)$\tcp*{Median extraction}
    $r=floor((N+1)/2)$;\\
    $indx(i)=I(r)$;\\
    $i=i+1$\tcp*{Deterministic domain}
    $w\leftarrow tag /sum(tag)$;\\ 
    $q\leftarrow cumsum(w)$\\
    \While{i$<=$N}{
    $sampl \leftarrow rand$;\\
    $j \leftarrow 1$;\\

\While{q(j)$<$sampl}{
 $j \leftarrow j+1$\tcp*{Update the tag}

}
$tag(i)\leftarrow tag(j)$;\\
$i \leftarrow i+1$;\\

 }


\end{algorithm}

\subsection{ Theoretical Asymptotic Behavior of Approximations}

\subsubsection{Central limit theorem}

Suppose that for each $t \in[1,T]$, $\tilde{X}_{t}^{(1)},...,\tilde{X}_{t}^{(M)}$ are independent, where $\tilde{X}_{t}^{(m')}, m'\in [1,M]$ denotes the median of the originator particles. For others $\tilde{X}_{t}^{(i)}, i \neq m'$ belong to the deterministic domain; the probability space of the sequence recursively changes with $t$ for sequential Monte Carlo, such a collection is called a triangular array of particles. Let $S_m:=\tilde{X}_{t}^{(1)}+...+\tilde{X}_{t}^{(M)}$. We expand the characteristic function of each $\tilde{X}_t^{(i)}$ to second-order terms and estimate the remainder to establish the asymptotic normality of $S_m$. Suppose that both the means and the variance are finite; {\color{red}we h}ave
\begin{equation}
\label{resample34}
E(X_t^{(i)})=\int \Psi(x)\pi (x)dx,\delta_{t,i}^2(\Psi )=E[(X_t^{(i)}-E(X_t^{(i)}))^2].
\end{equation}
{\bf{Theorem 1}} For each $t$ the sequence $\tilde{X}_{t}^{(1)},...,\tilde{X}_{t}^{(M)}$ sampled from the originator particles  $X_{t}^{(1)},...,X_{t}^{(N)}$ , suppose that are independent, where $\tilde{X}_{t}^{(m')}, m'\in [1,M]$ denotes the median of the originator particles. For the rest $\tilde{X}_{t}^{(i)}, i \neq m'$ belong to the deterministic domain;  
let $\Psi$ be a measurable function and assume that there exists $\bf{\tilde{X}_t} \subset \mathfrak{K}$ satisfying
\begin{equation}
\label{resamplet7}
\int_{x\in \mathfrak{K}}^{}\pi(dx)\mathbb{E}_x\left[\sum_{t=1}^{T}\left|\Psi(\bf{X_t})\right|^{2+\epsilon}\right]< \infty
\end{equation}
and
\begin{equation}
\label{resamplef2}
\text{sup}_{x\in \mathfrak{K}}\mathbb{E}_x\left[\sum_{t=1}^{T}\left|\Psi(\bf{X_t})\right|\right]< \infty,
\mathbb{E}_{\pi_t}[\Psi]:=\int_\mathfrak{K}\pi(dx)\mathbb{E}_x\left[\sum_{i=1}^{N}\Psi(\bf{X^{(i)}})\right]<\infty.
\end{equation}
 If $\bf{\tilde{X}_t}$ is aperiodic, irreducible, positive Harris recurrent with invariant distribution $\pi$ and geometrically ergodic, and if, in addition, 
\begin{equation}
\label{resample3fr}
\delta_{t,i}^2(\Psi ):=\int\pi(dx)\mathbb{E}_x\left[\left(\Psi({\tilde{X}_t^{(i)}})-\mathbb{E}_{\pi_t}[\Psi]\right)^2\right]< \infty,s_m^2= \lim_{M \to \infty}\sum_{i=1}^{M}\delta_{t,i}^2(\Psi ),
\end{equation}

$\{\Psi({\tilde{X}_t^{(i)}}\}$ satisfies

\begin{equation}
\label{resample3f}
\lim_{M \to \infty}\sum_{i=1}^{M}\left\{\Psi({{\tilde{X}_t^{(i)}}})-\mathbb{E}_{\pi_t}[\Psi]\right\}
\sim N(0,s_m^2).
\end{equation}


{\bf{Proof}} Let $Y_{t,i}=\Psi({{\tilde{X}_t^{(i)}}})-\mathbb{E}_{\pi_t}[\Psi]$, 
by \cite{billingsley1995measure}, $\left| e^{iy}-\sum_{k=0}^{M}\frac{(iy)^k}{k!} \right|\leq \min\{\frac{(y)^{M+1}}{(M+1)!},\frac{2(y)^M}{M!}\}$, when $M=2$, we have
\begin{equation}
\label{resamplee4}
\left| e^{iy}-(1+iy-\frac{1}{2}y^2) \right|\leq \min\{\frac{1}{6}\left|y\right|^3,\left|y\right|^2\}.
\end{equation}
We first assume that $\Psi(\cdot)$ is bounded. From the property of characteristic function, the left-hand side can be written as
$\left| \mathbb{E}\left[e^{(i\lambda Y_{t,i})}|\mathfrak{K}  \right]-(1-\frac{\lambda^2\delta_{t,i}^2(\Psi)}{2})\right|$

Therefore, the corresponding character function $\varphi_{t,i}(\lambda)$ of $Y_{t,i}$ satisfies
\begin{equation}
\label{resampleV1}
\left| \varphi_{t,i}(\lambda)-(1-\frac{\lambda^2\delta_{t,i}^2(\Psi)}{2}) \right|\leq \mathbb{E}\left[ min\{\left|\lambda Y_{t,i}\right|^2,\frac{1}{6}\left|\lambda Y_{t,i}\right|^3\}\right].
\end{equation}

Note that the expected value exists and is finite, the right-hand side term can be integrated by
\begin{equation}
\label{resamplew2}
\int_{\left|Y_{t,i}\right|\geq \epsilon \delta_{t,i}  \sqrt{M}}  \min\{\left|\lambda Y_{t,i}\right|^2,\frac{1}{6}\left|\lambda Y_{t,i}\right|^3\}dx
\end{equation}

As $M \to  +\infty$, $\{Y_{t,i}\} \to  \emptyset $, then, $E\left[ min\{\left|\lambda Y_{t,i}\right|^2,\frac{1}{6}\left|\lambda Y_{t,i}\right|^3\}\right] \to 0$, which satisfies Lindeberg condition:
\begin{equation}
\label{resampler3}
\lim_{M \to \infty}\sum_{i=1}^{M} \frac{1}{s_n^2}\int_{{\left|Y_{t,i}\right| \geq \epsilon \delta_{t,i} \sqrt{M}}}^{}Y_{t,i}^{2}dX=0
\end{equation}
for $\epsilon>0, s_m^2= \sum_{i=1}^{M}\delta_{t,i}^2(\Psi )$.
\begin{equation}
\label{resamplet5}
\lim_{M \to \infty}\left| \varphi_{t,i}(\lambda)-(1-\frac{\lambda^2\delta_{t,i}^2(\Psi)}{2}) \right|=0.
\end{equation}

By \cite{billingsley1995measure} 
\begin{equation}
\label{resamplw3j}
 \varphi_{t,i}(\lambda)=1+i\lambda \mathbb{E}[X]-\frac{1}{2}\lambda^2\mathbb{E}[X^2]+o(\lambda^2),\lambda \to 0.
 \end{equation}

By page358 Lemma 1 \cite{billingsley1995measure}.

\begin{align}
\label{eqn:eqlabel20d3j}
\begin{split}
& \left| \prod_{i=1}^{M}e^{-\lambda^2\delta_{t,i}^2(\Psi)/2}-\prod_{i=1}^{M}(1-\frac{1}{2}\lambda^2\delta_{t,i}^2(\Psi)) \right|\leq \sum_{i=1}^{M}\left| e^{-\lambda^2\delta_{t,i}^2(\Psi)/2}-1+ \frac{1}{2}\lambda^2\delta_{t,i}^2(\Psi)\right|\leq\\
&\sum_{i=1}^{M} \left[ \frac{1}{4}\lambda^4\delta_{t,i}^4(\Psi) \sum_{j=2}^{\infty} \frac{\frac{1}{2^{j-2}}\lambda^{2j-4}\delta_{t,i}^{2j-4}(\Psi)}{j!}    \right]\leq  \sum_{i=1}^{M} \frac{1}{4}\lambda^4\delta_{t,i}^4(\Psi)e^{\left|\frac{1}{2}\lambda^2\delta_{t,i}^2 (\Psi)\right|}
\end{split}
\end{align}

Thus,
\begin{equation}
\label{resamplw4t}
 \prod_{i=1}^{M}e^{-\lambda^2\delta_{t,i}^2(\Psi)/2}=\prod_{i=1}^{M}(1-\frac{1}{2}\lambda^2\delta_{t,i}^2(\Psi))+o(\lambda^2)=\prod_{i=1}^{M}e^{-\lambda^2\delta_{t,i}^2(\Psi)/2}+o(\lambda^2)=e^{-\frac{\lambda^2s_m^2}{2}}+o(\lambda^2)
 \end{equation}

The characteristic function $\prod_{i=1}^{M}\varphi_{t,i}(\lambda)$ of $\sum_{i=1}^{M}Y_{t,i}=\sum_{i=1}^{M}\left\{\Psi({\bf{X_t^{(i)}}})-\mathbb{E}_{\pi_t}[\Psi]\right\}$ is equal to $e^{-\frac{\lambda^2s_m^2}{2}}$, thus,
~\eqref{resample3f} holds.

\subsubsection{Asymptotic Variance 1}

The sample median can be defined as
\begin{equation}\label{eqn:eqlabesded}
X_t^{(m')}=
\begin{cases}
 &X_t^{(\frac{1}{2}(1+N))}, N=2r+1, r \in \mathbb{R^+} \\ 
 &\frac{1}{2}(X_t^{(\frac{1}{2}N)}+X_t^{(\frac{1}{2}N+1)}), N=2r, r \in \mathbb{R^+} . 
 \end{cases}
\end{equation}

Define 
\begin{align}
\label{eqn:eqlabeler3}
\begin{split}
  \mathbb{E}_{q_{k,t+1}}({\Psi}\mid {\bf{X_{t}}})
 = \begin{cases}
 & \int {\Psi}({\bf{X_{t}}})\prod_{l=t+1}^{k}q_{l}({\bf{X_{l}}}\mid {\bf{X_{l-1}}})  {\bf{X_{l}}}, \text{if} \ \ t<k,\ \\
 & q({\bf{X_{k}}}) \ \  \text{otherwise}. 
\end{cases}
\end{split}
\end{align}

and
\begin{equation}
\label{resaplfj4}
w_{ts}=\frac{\pi_t(X_t \mid X_{t-1},Y_{1:t})}{\pi_s(X_s\mid X_{s-1},Y_{1:s})\prod_{l=s+1}^{t}q_l(X_l\mid X_{l-1,Y_{1:l}})}.
\end{equation}



{\bf{Theorem 2}} Under the integrability conditions of theorem 1, suppose that $\tilde{X}_{t}^{(i)} \to U[0,1]$, $\lim_{M \to \infty}\sum_{i=1}^{M}\left\{\Psi({\bf{\tilde{X}_t^{(i)}}})-\mathbb{E}_{\pi_t}[\Psi]\right\}
\sim N(0,s_m^2)$, satisfying
\begin{equation}
\label{resaplfsd24}
s_m^2\leq (M-1) \cdot V_{t,t_0}({\Psi})+\frac{1}{8r+12},
\end{equation}
 where the originator particle size $N=2r+1, V_{t,t_0}({\Psi})=\frac{1}{M}\sum_{s=t_0}^{t}\mathbb{E}_{\pi_s}\mathbb{E}_{q_{s+1}}[\mathbb{E}_{q_{s+2}}\cdots \mathbb{E}_{q_{t}}\{({\Psi}-\mathbb{E}_{\pi_t}({\Psi}))w_{ts}\}]^2$ and
\begin{equation}
\label{resaplfgy}
V_{t,t_0}({\Psi})>\frac{1}{M}\sum_{s=t_0}^{t}\mathbb{E}_{q_{t}}\left[({\Psi}-\mathbb{E}_{\pi_t}({\Psi}))w_{ts}\right]^2.
\end{equation}

{\bf{Proof}} We decompose the original sequence $\tilde{X}_{t}^{(1)},...,\tilde{X}_{t}^{(M)}$ in descending order into two parts, $\tilde{X}_t^{(m')}$ denotes the median of the originator particles, the rest $\tilde{X}_{t}^{(i)}, i \neq m'$ belong to the deterministic domain; We solve for the variance of these estimators separately.

We assume that the population has an infinite number of individuals,  The values of the
 variance of the median for $2r$ even and $2r+1$ odd approach the same limit,
 but the value for the even will be less than the value for the odd \cite{hojo1931distribution}, Karl Pearson extended it with a more accurate estimation of the variance. Consequently, for the upper bound, here we consider the variance at the case of $N=2r+1$, denoted by $V({\Psi}'), {\Psi}'={\Psi(\tilde{X}_t^{\frac{1}{2}(1+N)})}$. Next, we derive a more detailed expression separately.

For $V({\Psi}')$, we first need to find the pdf of $\tilde{X}_t^{r+1}$, intuitively,
\begin{equation}
\label{resampleg2}
\mathbb{P}(\tilde{X}_t^{1+r}\in dx)=\sum_{i=1}^{2r+1}\mathbb{P}(\tilde{X}_t^{i}\in dx,B_i)
\end{equation}
where $B_i$ is the event that $r$ of the $2r$ values $V_1,...,V_{i-1},V_{i+1},...,V_{2r+1}$ are less than $x$. Because $V_i$ does not appear in the event $B_i$, the event $\{V_i \in dx\}$ is the chance that if we toss a coin $2r$ times, the probability of $r$ tails obtained, it can be formulated as
\begin{equation}
\label{resamplf32}
\mathbb{P}(\tilde{X}_t^{1+r}\in dx)=(2r+1)\cdot \mathrm{C}_{2r}^{r}x^r(1-x)^rdx=(2r+1)\binom{2r}{r}x^r(1-x)^rdx=\frac{(2r+1)!}{2r!}x^r(1-x)^rdx.
\end{equation}

Thus, ${\Psi}'\sim Beta(r+1,r+1)$, The variance of ${\Psi}'$ is $V({\Psi}')=\frac{1}{8r+12}$.

The particle in the deterministic domain  was resampled on the basis of the originator particle, which has been truncated, satisfying
\begin{equation}
\label{resaplg2}
\tilde{X}_{0:t}^{(i)}={X}_{0:t}^{(i)} \cdot {\mathbbm{1}}\left[\tilde w_t^{i} \geq \frac{1}{N}\right], i \neq m'.
\end{equation}

For any generic function ${\Psi}({X}_{0:t}^{(i)})$, the corresponding sample mean after resampling
\begin{equation}
\label{resapltd2}
\bar{\Psi}_t=\frac{1}{N}\sum_{i=1}^{N}{\Psi}_t(\tilde{X}_t^{(i)})
\end{equation}

is a consistent estimator of $\mathbb{E}_{\pi_s}[{\Psi}]$, whose variance $V_{t,t_0}({\Psi})$ is a function of the incremental weights and transition kernels encountered up to time $t$ from $t_0$. Inspired by \cite{berzuini1997dynamic}, the variance of this estimator  under large sample sizes can be formulated as

\begin{equation}
\label{resapldg2}
V_{t,t_0}({\Psi})=\frac{1}{M}\sum_{s=t_0}^{t}\mathbb{E}_{\pi_s}\mathbb{E}_{q_{s+1}}[\mathbb{E}_{q_{s+2}}\cdots \mathbb{E}_{q_{t}}\{({\Psi}-\mathbb{E}_{\pi_t}({\Psi}))w_{ts}\}]^2
\end{equation}

where $\mathbb{E}_{\pi_s}$ denotes expectation under the posterior distribution $\pi (X_{0:s} \mid y_{1:s})$, $\mathbb{E}_{q_{s+1}}$ denotes expectation under importance sampling distribution $q_{s+1}(X_{s+1} \mid X_{0:s},y_{1:s+1})$, and $N_s$ is the population size at $t=s$.

Consequently, each resampling stage at $s$ contributes an additional variance component ~\eqref{resapldg2}.

After rejuvenation, many of the previous particles will be discarded, and if we assume the whole population size of particles is stable, then the limit on the proportion of discarded particles satisfies,
\begin{equation}
\label{resapg34}
\lim_{t \to \infty,N \to \infty} \left( 1-\frac{\mathbbm{1}\left[\tilde w_t^i \geq \frac{1}{N}\right]}{\sum_{j=1}^{n}\tilde w_j} \right)^t\approx \left( 1-\frac{\epsilon}{N} \right)^t\approx e^{-\epsilon}
\end{equation}

Although the progressive impoverishment will lead to an increase in variance $V_{t,t_0}$, the rest will maintain a common attribute $\tilde w_t^i > \frac{1}{N}$ when $n \to \infty$. the accumulation of variance components after each rejuvenation will be negligible. 

As the simulation consistent estimator of $V_t$ is not available from the output samples, we consider the case from importance sampling distribution $q(X_{s},dX_{s+1})$ of $X_{s+1}=X_{s}$, the variance can be reduced to 

\begin{equation}
\label{resapldg2ty}
V'_{t,t_0}({\Psi})=\frac{1}{M}\sum_{s=t_0}^{t}\mathbb{E}_{\pi_s}\mathbb{E}_{q_{s+1}}[\mathbb{E}_{q_{s+2}}\cdots \mathbb{E}_{q_{t}}\{({\Psi}-\mathbb{E}_{\pi_t}({\Psi}))w_{ts}\}]^2=\frac{1}{M}\sum_{s=t_0}^{t}\mathbb{E}_{q_{t}}\left[({\Psi}-\mathbb{E}_{\pi_t}({\Psi}))w_{ts}\right]^2
\end{equation}

\begin{equation}
\label{resap3f4}
w_{ts}=\prod_{l=s+1}^{t}w_{l,l-1} \propto \prod_{l=s+1}^{t}W_{l}=\prod_{l=s+1}^{t}\frac{g(y_l \mid X_l^i)f(X_l^i \mid X_{l-1}^i)}{q(X_l^i\mid X_{0:l-1}^i,y_{1:l})}
\end{equation}

A simulation-consistent estimator of $V'_{t,t_0}$ is

\begin{equation}
\label{resapgo2}
\hat{V}'_{t,t_0}=\frac{1}{M}\sum_{s=t_0}^{t}\frac{\sum_{j=1}^{n_t}\{{\Psi}(\epsilon_t^{(j)})-\bar{\Psi}_t\}^2\prod_{l=s+1}^{t}W_l^{r(j)}}{\sum_{j=1}^{n_t}\prod_{l=s+1}^{t}W_l^{r(j)}}
\end{equation}

where $r(j)$ is the index of the sample at stage $r$ that survives as sample $j$ at stage $t$. $\hat{V}'_{t,t_0}$ provides an indicator of sample size $n_t$ whether it is adequate to resist particle impoverishment. Thus, ~\eqref{resaplfsd24} and  ~\eqref{resaplfgy} hold.

{\bf{Theorem 3}} Suppose that $\tilde{X}_{t}^{(1)},...,\tilde{X}_{t}^{(M)}$ each with a strictly positive probability density function and continuity on $\mathbb{R}$, let $m'_i$ be the median of each $\tilde{X}_{t}^{(i)}$, such that the cumulative function of $\tilde{X}_{t}^{(i)}$ satisfying $F(m'_i)=\frac{1}{2}$, then the sample median $\mathcal{M}$ of $\left\{ \tilde{X}_{t}^{(1)},...,\tilde{X}_{t}^{(M)} \right\}$ approximates the $\mathbb{N}(m'_i,\frac{1}{\sigma_0^2(m'_i)})$ distribution in the precise sense that, as $M \to \infty$,
\begin{equation}
\label{resa444pgo2}
\lim_{M \to \infty} \mathbb{P}\left[ \frac{\mathcal{M}-m'_i}{\sigma_0(m'_i)}\le x \right]=\Phi(x) \ \ \ (x\in \mathbb{R}),
\end{equation}
where $\sigma_0^2(m'_i)>4Mf(m'_i)^2$.

{\bf{Proof}} We follow ~\eqref{resampleg2} and let $x=\frac{1}{2}+\frac{1}{2}\frac{y}{\sqrt{2r}},dx=\frac{dy}{2\sqrt{2r}}$, we have
$\binom{2r}{r}\frac{1}{2^{2r}} \sim\frac{1}{\sqrt{\pi r}},$

\begin{equation}
\label{resgfh3}
\int_{-\infty}^{+\infty}\lim_{r \to \infty} (1-\frac{y^2}{2r})^rdy=\int_{-\infty}^{+\infty}e^{-\frac{y^2}{2}}dy=\sqrt{2\pi}
\end{equation}

As $r \to \infty$, combining ~\eqref{resampleg2} and ~\eqref{resgfh3}, we have
\begin{equation}
\label{resg33fh3}
\frac{(2r+1)!}{2r!}x^r(1-x)^rdx=\frac{1}{\sqrt{2\pi}}e^{-\frac{y^2}{2}}dy
\end{equation}
Thus, the quantity $\mathcal{M}$ can be expressed by $\mathcal{M}=\frac{1}{2}+\frac{Y}{2\sqrt{2r}}$, where $Y\sim \mathbb{N}(0,1)$.

Since $F(X) \in[0,1]$, $F$ is continous and strictly increasing on $\mathbb{R}$. $F(\mathcal{M})$ is the sample median of the $F(\tilde{X}_{t}^{(i)})$, from the Taylor series, it satisfies
\begin{equation}
\label{resgd33fh3}
F(\mathcal{M})=\frac{1}{2}+\frac{Y}{2\sqrt{2r}} > F(m'_i)+(\mathcal{M}-m'_i)f(m'_i).
\end{equation}
Since $F(m'_i)=\frac{1}{2}$, it yields to
\begin{equation}
\label{redsgdd3f}
\mathcal{M}-m'_i <\frac{Y}{2f(m'_i)\sqrt{M}}.
\end{equation}
Thus, $\sigma_0^2(m'_i)>4Mf(m'_i)^2$.
We have the same limiting value when $M=2r$ \cite{hojo1931distribution,williams2001weighing}.

\subsubsection{Consistency}

{\bf{Theorem 4}} Assume that the particle set $\left \{ X_{0:t}^{(i)},w_t^i \right \}, i\in [1,N] $ on the state space $ \Omega $ is consistent, where the convergence of Markovinian state transition holds. Then, the uniform weighted sample  $\left \{ \tilde{X}_{0:t}^{(i)},\tilde{w}_t^i \right \}, i\in [1,M] $ in the subset of  $ \Omega $ drew by the repetitive ergodicity in deterministic domain with median resampler is biased but consistent. 

{\bf{Proof}} There is a special case that when the median of particles belong to the deterministic domain, the particles with weight $\tilde{w}_t^i \leq \frac{1}{N}$ have been totally discarded, thus, the particle set after resampling is biased.

Under the integrability conditions of theorem 1, We invoke Chebyshev's Inequality,

$Y_{t,i}=\Psi({{\tilde{X}_t^{(i)}}})-\mathbb{E}_{\pi_t}[\Psi], i\neq m'$, 

\begin{equation}
\label{resapdg4}
V_{t,t_0}({Y_{t,i}})=V_{t,t_0}({\Psi})=\frac{1}{M}\sum_{s=t_0}^{t}\mathbb{E}_{\pi_s}\mathbb{E}_{q_{s+1}}[\mathbb{E}_{q_{s+2}}\cdots \mathbb{E}_{q_{t}}\{({\Psi}-\mathbb{E}_{\pi_t}({\Psi}))w_{ts}\}]^2
\end{equation}

As  $M \to \infty$, $V_{t,t_0}({Y_{t,i}}) \to 0$, for $V({\Psi}')=\frac{1}{8r+12}$, as $r \to \infty, V({\Psi}') \to 0$, consequently, 
\begin{equation}
\label{resapdfgg24}
\lim_{M\to \infty,r\to \infty}s_m^2\leq \lim_{M\to \infty,N\to \infty}\left[(M-1) \cdot V_{t,t_0}({\Psi})+\frac{1}{8r+12}\right]=0,
\end{equation}

$\mathbb{E}(Y_{t,i})^2=V(Y_{t,i})+[\mathbb{E}(Y_{t,i})^2]=0$

\begin{equation}
\label{reh6ygg24}
P(\left|Y_{t,i}\right|\leq \epsilon)=P(Y_{t,i}^2\geq \epsilon^2)\leq \frac{\mathbb{E}(Y_{t,i})^2}{\epsilon^2}=0.
\end{equation}

$\lim_{n \to \infty} P(\left|Y_{t,i}\right|\leq \epsilon)=0$ for all $\epsilon \leq 0.$ $Y_{t,i}$ is consistent.

As the resampling schema repetitively in a scaled domain, the total variance of our method obtained will be the lowest compared to other resampling methods, which is verified by the experiments.

\section{Simulation}

In this part, the results of the comparison of these resampling methods are validated from the experiments with the linear Gaussian state space model and nonlinear state space model, respectively. We ran the experiments on an HP Z200 workstation with an Intel Core i5 and an $\#82-18.04.1-$ Ubuntu SMP kernel. The code is available at \url{https://github.com/986876245/Variance-Reduction-for-SMC}.



 \subsection{Linear Gaussian State Space Model}


This linear model is expressed by:

\begin{equation}
X_0\sim \mu(X_0),\ \ \
X_t\mid X_{t-1} \sim N(X_t;\phi X_{t-1},\delta _v^2),\ \ \
Y_t\mid X_t \sim N(y_t;X_t,\delta _e^2).
\end{equation}

\begin{figure}[!htp]
	\centering
	\begin{minipage}{.49\textwidth}
    
  \captionsetup{justification=centering,margin=0.5cm}
		\centering
		\includegraphics[width=.95\linewidth]{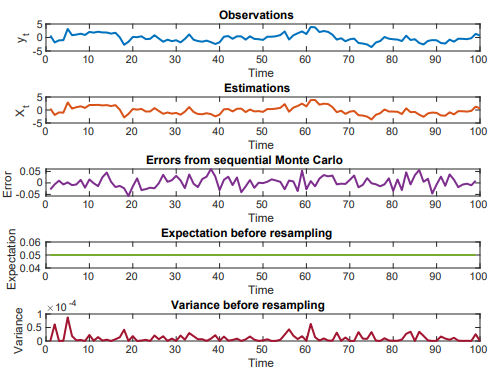}
		\captionof{figure}{Repetitive deterministic traverse resampling for linear Gaussian State Space Model.}
		\label{fig:experiment1}
	\end{minipage}
	\begin{minipage}{.49\textwidth}
		\centering
		\captionsetup{justification=centering,margin=0.1cm}
		\includegraphics[width=.95\linewidth]{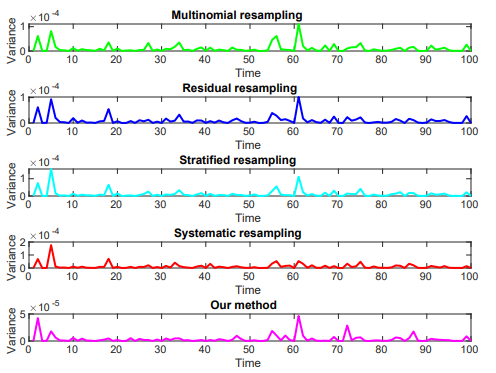}
		\captionof{figure}{Variance analysis for different resampling strategies.}
		\label{fig:experiment3}
	\end{minipage}%
\end{figure}

We keep parameters the same as \cite{dahlin2015getting} to compare with the different resampling
methods. Where $\theta =\left \{ \phi,\delta_v,\delta_e \right \}$,$\phi \in (-1,1)$ describes the
persistence of the state, while $\delta_v,\delta_e$ denote the standard deviations of the state
transition noise and the observation noise, respectively. The Gaussian density is denoted by
$N(x;\mu ,\delta ^2)$ with mean $\mu $ and standard deviation $\delta>0$. In
\autoref{fig:experiment1}, we use 20 particles to track the probability distribution of the state,
composed of 100 different times, the ground truth is from the Kalman filter
\cite{welch1995introduction}, 
the error denotes the difference between the estimation by SMC and the ground truth. Initially, the
expectation of weights for 20 particles is equal to $\frac{1}{20}$, which means that these particles
have equal functions to track the state.  

For the resampling procedure, we compare the variance from different classical resampling methods, shown in \autoref{fig:experiment3}. The variance from the deterministic traverse method is the smallest. Thus, the effective particles are more concentrated after resampling based on our proposal.


\begin{figure}[!htp]
	\centering
	\begin{minipage}{.49\textwidth}
		\captionsetup{justification=centering,margin=0.5cm}
		\centering
		\includegraphics[width=.95\linewidth]{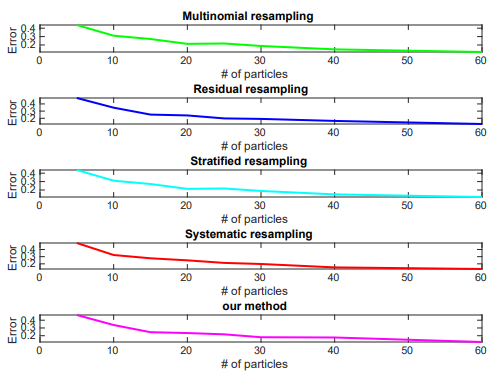}
		\captionof{figure}{RMSE analysis for different resampling strategies.}
		\label{fig:experiment4}
	\end{minipage}
	\begin{minipage}{.49\textwidth}
		\centering
		\captionsetup{justification=centering,margin=0.1cm}
		\includegraphics[width=.95\linewidth]{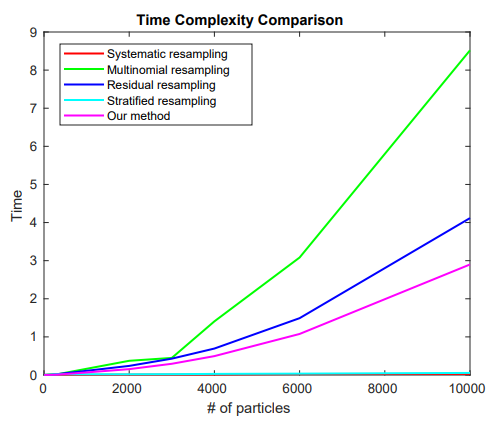}
		\captionof{figure}{Time complexity analysis for different resampling strategies.}
		\label{fig:experiment5}
	\end{minipage}%
\end{figure}

\autoref{fig:experiment4} shows the root mean squared(RMSE) error for different resampling strategies, the decreasing rate of our method is higher than that from other methods as the particle increase, given in a feasible domain. 

The computational complexity is another factor the resampling algorithms are compared on, \autoref{fig:experiment5} shows the execution times for different particles distributed, generally, it depends on the machines and random generator, during our simulations, the time consumption is different under the same condition of resampling method and number of particles. Furthermore, we find under the same resampling methods, the time consumed for the small size of particles is much more than that of the larger ones. The computational stability of particles with resampling methods is very sensitive to the units from a specific population. For safety, we conduct multiple experiments to achieve the general complexity trend. 


\begin{figure}[!htp]
	\centering
	\begin{minipage}{.49\textwidth}
		\centering
		\captionsetup{justification=centering,margin=0.1cm}
		\includegraphics[width=.95\linewidth]{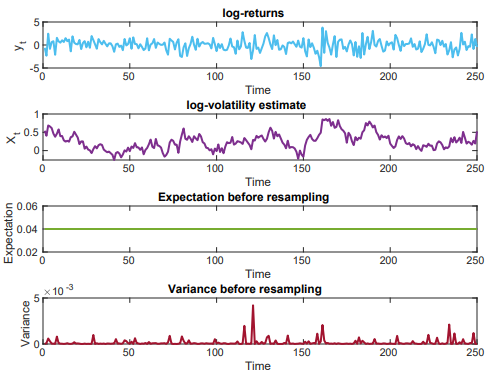}
		\captionof{figure}{(a) The daily log-returns. (b) The estimated log-volatility with 95\% confidence intervals of the NASDAQ OMXS30 index for the period from January 2,2015 to January 2,2016. (c) The expectation of weights for particles before resampling. (d)The variance of weights for particles before resampling.}
		\label{fig:experiment21}
	\end{minipage}%
	\begin{minipage}{.49\textwidth}
		\centering
		\captionsetup{justification=centering,margin=0.1cm}
		\includegraphics[width=.95\linewidth]{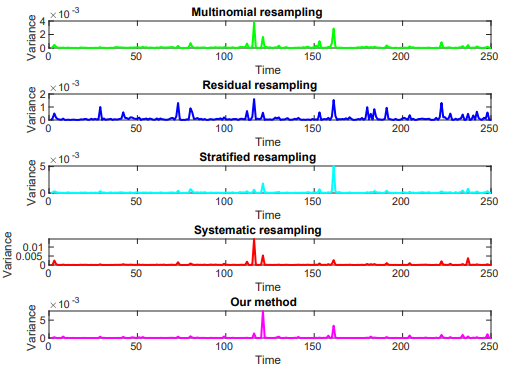}
		\captionof{figure}{Variance analysis of different resampling strategies for nonlinear state space model.}
		\label{fig:experiment23}
	\end{minipage}%
\end{figure}

In \autoref{fig:experiment5}, all the experiments are conducted under the same conditions, for large-size particles, the stratified and systematic strategies are favorable. In  Table~\ref{tab:data}, we can find under small-size particles(less than 150), our method performs best.

\renewcommand{\arraystretch}{1.5} 
\begin{table*}[ht]  
  \centering  
  \fontsize{6.5}{8}\selectfont  
  \caption{Time complexity analysis of different resampling strategies.} \label{tab:data} 
  \label{tab:performance_comparison}  
    \begin{tabular}{cccccc }  
    \toprule  
    {\# of particles}&Multinomial resampling&Residual resampling&Systematic resampling&Stratified resampling&{\bf Our method}\cr
    
      \hline
    5&$0.0026$&$0.0021$&$0.0019$&$0.0023$&{\bf {0.0016}}\cr  
    15&$0.0036$&$0.0025$&$0.0022$&$0.0033$&{\bf {0.0018}}\cr 
    50&$0.0057$&$0.0027$&$0.0024$&$0.0047$&{\bf {0.0022}}\cr  
    80&$0.0087$&$0.0032$&$0.0029$&$0.0051$&{\bf {0.0027}}\cr
    100&$0.0127$&$0.0038$&$0.0034$&$0.0067$&{\bf {0.0030}}\cr
    150&$0.0161$&$0.0043$&$0.0038$&$0.0088$&{\bf {0.0036}}\cr  
    \bottomrule  
    \end{tabular}  
\end{table*}

\subsection{Nonlinear State Space Model}
We continue with a real application of our proposal to track the stochastic volatility, a nonlinear State Space Model with Gaussian noise, where log volatility considered as the latent variable is an essential element in the analysis of financial risk management. 
The stochastic volatility is given by
\begin{equation}
X_0\sim N(\mu,\frac{\sigma _v^2}{1-\rho  ^2}),
X_t\mid X_{t-1} \sim N(\mu+\rho  (X_{t-1}-\mu),\sigma _v^2),
Y_t\mid X_t \sim N(0,exp(X_t)\tau).
\end{equation}
where the parameters $\theta =\left \{\mu, \rho  ,\sigma_v,\tau \right \}$, $\mu\in \mathbb{R},
\rho  \in [-1,1]$, $\sigma _v $ and $\tau \in \mathbb{R}_+$, denote the mean value,
the persistence in volatility, the standard deviation of the state process and the instantaneous
volatility, respectively. 

The observations $y_t=\log(p_t/p_{t-1})$,  also called log-returns, denote the logarithm of the daily difference in the exchange rate $p_t$, 
here, $\{{p_t}\}_{t=1}^T$ is the daily closing price of the NASDAQ OMXS30 index (a weighted average of the 30 most traded stocks at the Stockholm stock exchange). We extract the data from {\color{blue} \href{https://www.quandl.com/}{Quandl}} for the period between January 2, 2015 and January 2, 2016. The resulting log-returns are shown in \autoref{fig:experiment21}. We use SMC to track the time-series persistency volatility,  large variations are frequent, which is well-known as volatility clustering in finance, from the equation (42), as $|\phi|$ is close to $1$ and the standard variance is small, the volatility clustering effect easier occurs. We keep the same parameters as \cite{dahlin2015getting},
where $\mu \sim N(0,1), \phi \sim TN_{[-1,1]}(0.95,0.05^2)$, $ \delta _v \sim \hbox{\it Gamma}(2,10) $, $\tau=1$.

We use 25 particles to track the persistency volatility, the expectation of weights of particles is $\frac{1}{25}$, shown in \autoref{fig:experiment21}, it is stable as the same with \autoref{fig:experiment1}, the variance is in $10^{-3}$ orders of magnitude under random sampling mechanism.

In \autoref{fig:experiment23}, the variance from our proposal shows the minimum value at different times, nearly all the plot share the common multimodal feature at the same time, it stems from the multinomial distribution that both of them have when they resample a new unit.   

\section{Conclusion}

Resampling strategies are effective in Sequential Monte Carlo as the weighted particles tend
to degenerate. However,
we find that the resampling also leads to a loss of diversity among the particles.
This arises because in the resampling
stage, the samples are drawn from a discrete multinomial distribution, not a continuous one.
Therefore, the new samples
fail to be drawn as a type that has never occurred but stems from the existing samples 
by the repetitive schema. We have presented
a repetitive deterministic domain traversal for resampling and have achieved the lowest
variances compared to other resampling methods. As the size of the
deterministic domain $M\ll N$ (the size of population), our algorithm is faster than the state of the art, given a feasible size of particles, which is verified by theoretical deduction and
experiments of the hidden Markov model in both the linear and the non-linear case.

The broader impact of this work is that it can speed up existing sequential Monte Carlo applications
and allow more precise to estimates their objectives.
There are no negative societal impacts, other than those arising from the
sequential Monte Carlo applications themselves.

\section*{Acknowledgments}
This was supported in part by BRBytes project.

\bibliographystyle{unsrt}  
\bibliography{references}

\end{document}